# A Novel Method for Calculating Demand Not Served for Transmission Expansion Planning

Neeraj Gupta[1], *Student Member, IEEE*, Rajiv Shekhar[2], and Prem Kumar Kalra[1]

*Abstract*-- Restructuring of the power market introduced demand uncertainty in transmission expansion planning (*TEP*), which in turn also requires an accurate estimation of demand not served (*DNS*). Unfortunately, the graph theory based minimum-cut maximum-flow (*MCMF*) approach does not ensure that electrical laws are followed. Nor can it be used for calculating *DNS* at individual buses. In this letter, we propose a generalized load flow based methodology for calculating *DNS*. This procedure is able to calculate simultaneously generation not served (*GNS*) and wheeling loss (*WL*). Importantly, the procedure is able to incorporate the effect of $I^2R$ losses, excluded in *MCMF* approach. Case study on a 5-bus *IEEE* system shows the effectiveness of the proposed approach over existing method.

*Index Terms*—Graph theory, load flow analysis, power system reliability, planning, transmission lines.

## I. INTRODUCTION

Restructuring of the power market introduced demand uncertainty in transmission expansion planning [1]. Here, new lines are identified based on the minimization of the sum of the cost of the expected demand not served (E*DNS*) and capital cost for setting up additional transmission capacities [2], [3]. Consequently an accurate estimation of *DNS* is required. Unfortunately, the graph theory based *MCMF* approach does not ensure that electrical laws are followed [2] - [5]. Nor can it be used for calculating *DNS* at individual buses, referred to as *DNS$_s$*, an important input for "local" planning. In this article, we propose a generalized load flow based methodology to overcome the shortcomings of the *MCMF* approach. In fact, this methodology can also be used to simultaneously calculate *GNS* and *WL*. Here *GNS*/*DNS* can be calculated even while considering $I^2R$ losses, hereafter referred to as network losses (*NL*). The *MCMF* approach cannot incorporate the effect of network losses. In addition to minimizing *DNS*, the planner can minimize *WL* to improve the economics of the power systems.

## II. DNS

### A. Minimum-cut Maximum-flow Methodology

Using graph theory the *DNS* for the system is calculated according to the formula [2]:

P. K. Kalra and N. Gupta is with Electrical Engineering Department, Indian Institute of Technology Kanpur, UP, 208016, India (e-mail: kalra, ngtaj@iitk.ac.in).

R. Shekhar is with the Department of Materials Science and Engineering, Indian Institute of Technology Kanpur, UP, 208016, India (e-mail: vidtan@iitk.ac.in).

$$DNS = \sum_{s=1}^{b} D_s - \max(f_{S-L}) \quad (1)$$

$$\max(f_{S-L}) = \min_Q \left[ c(q_{S-L}) \right] \quad (2)$$

Here, $D_s$ represents the demand at bus 's', and $f_{S-L}$ is the flow from node S to node L. In equation (2), Q refers to the set of all such $q_{S-L}$ cuts, where $q_{S-L}$ *is* a set of elements whose removal from the graph breaks all directed paths from node S to node L. $c(q_{S-L})$ is the sum of the capacities of all the elements defining the $q_{S-L}$ cut. A graph theory representation of the network is given in section III.

### B. Proposed Methodology

Fig. 1 is a schematic diagram of a bus *"s"* in an electrical network with "*m*" incoming and "*n*" outgoing transmission lines. As a first step, an economic load dispatch using *DC-*load flow is run -- without considering transmission capacity constraints -- with the specified demand and generation at each node. *DNS* and *GNS* are then calculated at each bus *"s"* using the following relationships:

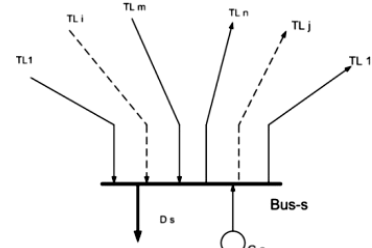

Fig. 1. Schematic diagram of a bus 's' in a network with '*m*' incoming and '*n*' outgoing transmission lines.

$$DIFF_s = D_s - \min\left(\sum_{i=1}^{m}\left(T_{f,i,s}, T_{c,i,s}\right)\right)$$
$$+ \min\left(\sum_{j=1}^{n}\left(T_{f,j,s}, T_{c,j,s}\right)\right) - G_s \quad (3)$$

$$DNS_s = DIFF_s : if\ DIFF_s > 0 \quad (4)$$

$$GNS_s = abs(DIFF_s) : if\ DIFF_s < 0 \quad (5)$$

$$DNS = \sum_{s=1}^{b} DNS_s \quad ; \quad GNS = \sum_{s=1}^{b} GNS_s \quad (6)$$

In fact, $DNS_s$/ $GNS_s$ calculations should only be made on buses that have at least one congested line, thereby reducing the computational time. It may be pointed out that $DNS \neq GNS$ in a network with network losses. *WL* can be calculated according to:

$$WL = \sum_{k=1}^{p} (T_{f,k,s} - T_{c,k,s}) \quad (7)$$

$G_s$ is generation at bus *'s'*. $T_{f,i,s}$ and $T_{c,i,s}$ refer to the actual power flow and transmission capacity of the $i^{th}$ incoming line to bus *'s'* respectively. Similarly *(i)* $T_{f,j,s}$ and $T_{c,j,s}$ and *(ii)* $T_{f,k,s}$ and $T_{c,k,s}$ have analogous meaning for the $j^{th}$ outgoing line from bus *'s'* and the $k^{th}$ congested line respectively.

### III. CASE STUDY

The base network used for the case study is shown in Fig. 2 [2]. Two additional cases have been generated by first changing the capacity of *(i) T1* from 100 MW to 25 MW (*case-2*) and *(ii) T2* from 75 MW to 25 MW (*case-3*). The graph theory representation of the power system with three cases is shown in Fig. 3 [2]. *S-1* and *S-2* branches represent the generation capacity at buses *1* and *2* respectively. Branches *2-L, 3-L, 4-L* and *5-L* represent the respective loads at buses *2* to *5*.

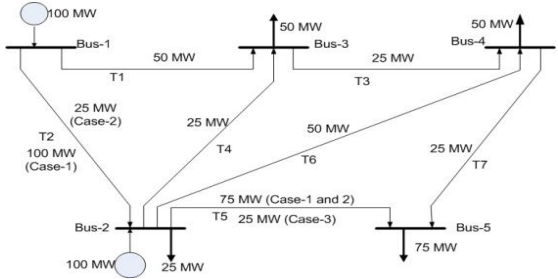

Fig. 2. Network for case study with specified generation, load and transmission capacities, generator at *bus-1* is slack generator.

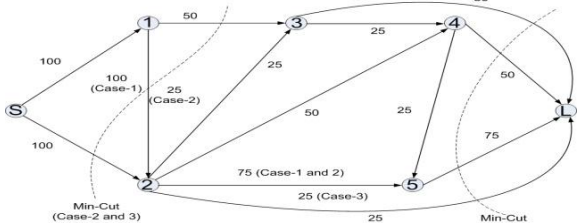

Fig. 3. Graph theory representation of Fig. 2 with given capacity of the branches.

Table-I shows *(i)* $DNS_s$ in all five buses and *(ii)* a comparison of *DNS* calculated by the proposed methodology (*PM*) and *MCMF* approach for a "lossless" network (*LLN*). Here the negative and positive values refer to *GNS* and *DNS* respectively. The shortcoming of the *MCMF* approach is illustrated by the results of cases *1* and *3*. In *case-1*, the *MCMF* approach suggests that the network is reliable and does not require any transmission expansion. On the other hand, *MCMF* underestimates the *DNS* by a factor of 2.4. Table-II compares the power flow in each transmission line using the proposed method and *MCMF* and shows the wheeling losses for the three cases.

A comparison of Tables-I and III shows that the presence of network losses changes the values of *(i) DNS* and *GNS* and *(ii) $DNS_s$/ $GNS_s$*. According to Table-III, It can be seen that the difference between *DNS* and *GNS* in cases-*1, 2* and *3* are 4.5 MW, 12.1 and 6 MW respectively, indicating more realistic situation. The *MCMF* approach cannot incorporate the effect of network losses.

TABLE I
*$DNS_s$/ $GNS_s$* FOR LOSSLESS NETWORK.

| Case | $DNS_s/GNS_s$ (MW) | | | | | *DNS* (MW) | |
|---|---|---|---|---|---|---|---|
| | Bus-1 | Bus-2 | Bus-3 | Bus-4 | Bus-5 | PM | *MCMF* |
| 1 | -24.9 | 0 | 9 | 15.9 | 0 | 24.9 | 0 |
| 2 | -25 | 0.1 | 9 | 15.9 | 0 | 25 | 25 |
| 3 | -25 | -36.2 | 9 | 15.9 | 36.4 | 61.3 | 25 |

TABLE II
POWER FLOW IN THE ELECTRICAL NETWORK

| | DC-Load Flow (*PM*) in MW | | | Graph Theory (*MCMF*) | | |
|---|---|---|---|---|---|---|
| | LLN | NL | | Flow | | |
| | | from | to | Case-1 | Case-2 | Case-3 |
| T1 | 74.85 | 77.94 | 73.56 | 50 | 50 | 50 |
| T2 | 25.15 | 28.83 | 28.68 | 50 | 25 | 25 |
| T3 | 40.83 | 40.19 | 40.03 | 25 | 25 | 25 |
| T4 | 15.98 | 16.81 | 16.62 | 25 | 25 | 25 |
| T5 | 61.39 | 63.17 | 61.73 | 75 | 25 | 25 |
| T6 | 22.78 | 23.70 | 23.38 | 50 | 50 | 50 |
| T7 | 13.61 | 13.40 | 13.27 | 25 | 25 | 25 |
| Maximum Flow at Minimum Cut (MW) | | | | 200 | 175 | 175 |
| *WL* using *PM* (MW), equation (7) | | | | 43.7 | 40.8 | 77.2 |

TABLE III
*$DNS_s/GNS_s$* FOR NETWORK WITH LOSSES

| Case | $DNS_s/GNS_s$ (MW) | | | | | *DNS* (MW) | *GNS* (MW) |
|---|---|---|---|---|---|---|---|
| | Bus-1 | Bus-2 | Bus-3 | Bus-4 | Bus-5 | | |
| 1 | -27.9 | 0 | 8.4 | 15 | 0 | 23.4 | 27.9 |
| 2 | -31.8 | -3.68 | 8.4 | 15 | 0 | 23.4 | 35.5 |
| 3 | -31.8 | -34.5 | 8.4 | 15 | 36.7 | 60.3 | 66.3 |

### IV. CONCLUSION

In this letter, we have quantitatively shown that the proposed methodology, which follows electrical laws, indeed yields *DNS/ GNS* which are different from those calculated using the *MCMF* approach. This approach is especially useful for TEP in countries such as India that do not follow the nodal pricing mechanism. The calculation of $DNS_s$ allows planners to prioritize the setting up of transmission lines based on the consumer mix (residential, commercial, industrial, agricultural etc.) in a particular bus. Moreover, calculation of $DNS_s/GNS_s$ can be used for locating and boosting or adding generation capacities.

References
[1] C. W. Lee, S. K. K. Ng, J. Zhong and F. F. Wu, "Transmission expansion planning from past to future", IEEE PES, Power Systems Conference and Exposition, 10.1109/PSCE, PSCE '06.
[2] C. T. Su and G. R. Lii, "Power system capacity expansion planning using Monte Carlo simulation and Genetic Algorithm", International Congress on Modelling and Simulation, Hobart, Australia, Proceedings, pp. 1450-1455, 1997.
[3] J. Choi, T. Tran, A. A. El-Keib, R. Thomas, H. Oh, and R. Billinton, "A method for transmission system expansion planning considering probabilistic reliability criteria ", IEEE Transaction on Power Systems, Vol. 20, No. 3, 2005.
[4] J. Choi, T. Tran, A. A. El-Keib, and J. Watada, "Transmission expansion planning considering ambiguities using fuzzy modeling", International Journal of Innovative Computing Information and Control, Vol. 4, No. 8, 2008.
[5] A. Dwivedi, Y. Xinghuo, P. Sokolowski, "Analyzing power network vulnerability with maximum flow based centrality approach" In Proc. 8[th] IEEE International Conference on industrial Informatics (INDIN), 2010.